\documentstyle[12pt,aasms4,epsf]{article}

\begin{document}

\title{Measuring the galaxy power spectrum with multiresolution
decomposition -- II. diagonal and off-diagonal power spectra of
the LCRS galaxies}

\author{XiaoHu Yang\altaffilmark{1,2}\quad Long-Long
Feng\altaffilmark{1,2,3}\quad YaoQuan Chu\altaffilmark{1,2}\quad
Li-Zhi Fang\altaffilmark{4}}

\altaffiltext{1}{Center for Astrophysics, University of Science
and Technology of China, Hefei, Anhui 230026,P.R.China}
\altaffiltext{2}{National Astronomical Observatories, Chinese
Academy of Science, Chao-Yang District, Beijing, 100012, P.R.
China} \altaffiltext{3}{Institute for Theoretical Physics, Academy
of Science, Beijing, 100080, P.R. China
}\altaffiltext{4}{Department of Physics, University of Arizona,
Tucson, AZ 85721}

\begin{abstract}

The power spectrum estimator based on the discrete wavelet
transform (DWT) for 3-dimensional samples has been studied. The
DWT estimator for higher than 1-dimensional samples provides two
types of spectra with respect to diagonal and off-diagonal modes,
respectively. The two types of modes have different spatial
invariance, and therefore, the diagonal and off-diagonal DWT power
spectra are very flexible to deal with configuration-related
problems in the power spectrum detection. With simulation samples
and the mock catalogues of the Las Campanas redshift survey
(LCRS), we show 1. the slice-like geometry of the LCRS doesn't
affect the off-diagonal power spectrum with ``slice-like'' mode;
2. the Poisson sampling with the LCRS selection function doesn't
cause more than 1-$\sigma$ error in the DWT power spectrum; and 3.
the powers of the mass or galaxy random velocity fluctuations,
which cause the redshift distortion, are approximately
scale-independent. These results insure that the uncertainties of
the power spectrum measurement are under control. The scatter of
the DWT power spectra of the six strips of the LCRS survey is
found to be rather small. It is less than 1-$\sigma$ of the cosmic
variance of mock sample in the wavenumber range $0.1 < k < 2$ h
Mpc$^{-1}$. To fit the detected LCRS diagonal DWT power spectrum
with CDM models, we find that the best-fitting redshift
distortion parameter $\beta$ is about the same as that obtained
from the Fourier power spectrum. The velocity dispersions $\sigma_v$
for models SCDM and $\Lambda$CDM are also consistent with other
$\sigma_v$ detections with the LCRS. A systematic difference between the
best-fitting parameters of diagonal and off-diagonal power spectra
have been significantly measured. This indicates that the
off-diagonal power spectra are capable of providing information about
the power spectrum of the galaxy velocity field.

\end{abstract}


\section{Introduction}

The power spectrum is one of most important statistical measure to
quantify the clustering features of large scale mass density
distribution traced by galaxies. It directly reflects the physical
scales of the processes that affect structure formation.

In paper I (Fang \& Feng 2000), the method of measuring galaxy
power spectrum with the multiresolution analysis of discrete
wavelet transformation (DWT) is developed. It has been shown that
the DWT analysis provides a lossless estimation of the power
spectrum. The DWT power spectrum estimator is optimized in the
sense that the spatial resolution automatically adapts to the
perturbation wavelength under consideration. Since the DWT
analysis has strong capability of suppressing the off-diagonal
components of the covariance of density fluctuations, the DWT
estimator is convenient to deal with data sets with a large
volume and complex geometry. Moreover, the DWT estimator can
fully avoid the alias effect which appears in usual binning
schemes. Because Poisson process possesses diagonal covariance in
the DWT representation, the Poisson sampling and selection
effects on the power spectrum are suppressed to a minimum.

Besides these technical advantages of the computation of power
spectrum, the DWT space-scale decomposition provides a physical
insight into the clustering behavior in phase space, i.e. scale plus
physical position. Even for second order statistics, i.e. the
covariance of density fluctuations, the space-scale decomposition
can reveal some clustering behaviors, such as two-point correlation
in phase space, which can not be measured by decomposition with modes
non-localized in phase space. Moreover, the covariance in
DWT representation can avoid the difficulty caused by the central
limit theorem (Fan \& Bardeen 1995.)

The algorithm developed in Paper I is mainly for 1-D sample. Many
algorithms for 1-D sample can directly be generalized into 3-D. A
problem especially for 3-D DWT power spectrum estimator is on
rotational invariance. 3-D DWT modes is given by the direct product
of 1-D DWT modes, and therefore, each of the 3-D DWT modes is
described by three positive integers $(j_1,j_2,j_3)$ corresponding
to the scales of the mode in $x_1$, $x_2$ and $x_3$ directions,
respectively.
Thus, there are two types of modes: diagonal mode with $j_1=j_2=j_3$;
off-diagonal mode for which the three numbers $(j_1,j_2,j_3)$ are not
the same. The DWT estimator can provide two types of power spectra:
1. diagonal power spectrum given by the powers on diagonal modes,
2. off-diagonal power spectrum given by the powers on off-diagonal
modes.

Because the two types of modes have different spatial invariance,
the diagonal and off-diagonal DWT power spectra are very flexible
to deal with configuration-related problems in the power spectrum
detection. In this paper, we will study both diagonal and
off-diagonal power spectra with the Las Campanas redshift survey
galaxies, and relevant simulation samples and mock catalogues. The
purpose is to demonstrate how the two types of power spectra are
applied to estimate the uncertainties of power spectrum
measurement caused by various anisotropies of the sample. It
includes, for instance, the irregular geometry of the survey with
very low filling factor, the scale-dependence of redshift
distortion, and the inhomogeneity caused by the selection
function. These results insure that the DWT estimator is capable
of providing a stable measurement of the LCRS galaxy power
spectrum and the best-fitting parameters of bias and velocity
dispersion.

The paper is organized as follows. In \S 2, the DWT power spectrum
estimator is briefly introduced with the emphasis on the property and
algorithm for the off-diagonal DWT power spectrum. \S 3 describes
the real and mock samples of galaxies. The applications of the
diagonal and off-diagonal power spectra to the
configuration-related problems are presented in \S 4. The results
of the DWT power spectra of the LCRS, both diagonal and off-diagonal,
and their error estimation are given in \S 5. The best-fitting values
of the bias parameters and velocity dispersion are given in \S 5
as well. Finally, conclusions are stated in \S 6.

\section{DWT power spectrum estimator}

\subsection{The DWT decomposition of galaxy distributions}

If the position measurement is perfectly precise, the observed
galaxy number density distribution can be written as
%eq1
\begin{equation}
n_g({\bf x})=\sum_{m=1}^{N_g}w_m\delta^D({\bf x-x}_m)
\end{equation}
where coordinate ${\bf x}=(x_1,x_2,x_3)$, $N_g$ is the total
number of galaxies, ${\bf x}_m$ is the position of the $m$th
galaxy, $w_m$ is its weight, and $\delta^D$ the 3-D Dirac $\delta$
function. Without loss of generality, we consider the sample which
is contained in a spatial box with spatial ranges of $x_1= 0-L_1$,
$x_2=0-L_2$ and $x_3= 0-L_3$. It is not necessary that the box
$L_1L_2L_3$ is fully filled by sample, i.e., allows $n_g({\bf
x})=0$ within some areas inside the box.

The observed galaxy number density distribution is considered to
be a realization of a Poisson point process with an intensity
$n({\bf x})=\bar{n}_g({\bf x})[1+\delta({\bf x})]$, where
$\bar{n}_g({\bf x})$ is selection function, and $\delta({\bf x})$
the density contrast fluctuation on the underlying matter field.
The goal of power spectrum measurement is to estimate the power
spectrum of the density fluctuations $\delta({\bf x})=[n({\bf
x})/\bar{n}({\bf x})]-1$ from the observed realization
$\delta_g({\bf x})=[n_g({\bf x})/\bar{n}_g({\bf x})]-1$.

Different power spectrum estimators adopt different decomposition
of $\delta_g({\bf x})$. The DWT decomposition is based on the
orthogonal and complete set of 3-D wavelet basis $\{\psi_{\bf
j,l}({\bf x})\}$, which can be constructed by a direct product
of 1-D wavelet basis as
%eq2
\begin{equation}
\psi_{\bf j,l}({\bf x})=\psi_{j_1,l_1}(x_1)\psi_{j_2,l_2}(x_2)
 \psi_{j_3,l_3}(x_3),
\end{equation}
where $j_i=0,1,2..$ ($i=1,2,3$) and $l_i=0...2^{j_i-1}$. The
wavelet $\psi_{\bf j,l}({\bf x})$ is localized in phase (scale and
physical position) space, i.e. it is non-zero mainly in a volume
$L_1/2^{j_1}\times L_2/2^{j_2}\times L_3/2^{j_3}$, and  around the
position $(x_1=l_1L_1/2^{j_1},
x_2=l_2L_2/2^{j_2},x_3=l_3L_3/2^{j_3})$ (Fang \& Thews 1998.) The
bases eq.(2) are orthonormal with respect both ${\bf j}$ and
${\bf l}$, i.e.
$$
\int \psi_{\bf j,l}({\bf x})\psi_{\bf j',l'}({\bf x})d{\bf x}
  =\delta^K_{\bf j,j'}\delta^K_{\bf l, l'},
$$
where $\delta^K$ is the Kronecker Delta function.

The DWT decomposition of $[n_g({\bf x})/\bar{n}_g({\bf x})]-1$ is
then given by
%eq3
\begin{equation}
\tilde{\epsilon}_{\bf j}^{\bf l}=\int \Bigl[\frac{n_g({\bf x})}
{\bar{n}_g({\bf x})}-1\Bigr]\psi_{\bf j,l}({\bf x})d{\bf x},
\end{equation}
Since wavelets are admissible, i.e. $\int \psi_{j,l}(x) dx=0$,
eq.(3) becomes
%eq4
\begin{equation}
\tilde{\epsilon}_{\bf j}^{\bf l}= \int \frac{n_g({\bf
x})}{\bar{n}_g({\bf x})}
      \psi_{\bf j,l}({\bf x})d{\bf x}.
\end{equation}
Thus the fluctuations in galaxy number density can be described by
the wavelet function coefficients (WFC) $\tilde{\epsilon}_{\bf
j}^{\bf l}$, which is the density fluctuation on scale ${\bf
j}=\{j_1,j_2,j_3\}$ localized at the position ${\bf
l}=\{l_1,l_2,l_3\}$.

\subsection{DWT power spectrum estimator}

For a given scale ${\bf j}$, there are totally $2^{j_1+j_2+j_3}$
WFC variables. All second order statistics on the scale ${\bf j}$
can be drawn from the covariance of these WFCs
%eq5
\begin{equation}
C_{\bf j}^{\bf ll'}=\tilde\epsilon_{\bf j}^{\bf l}
 \tilde\epsilon_{\bf j}^{\bf l'}.
\end{equation}
It has been shown in Paper I that the power of the fluctuations on
the modes with the scale index ${\bf j}$ can be estimated by
%eq6
\begin{equation}
P_{\bf j}=I_{\bf j}^2-N_{\bf j}.
\end{equation}
where
%eq7
\begin{equation}
I_{\bf j}^2=\frac{1}{2^{j_1+j_2+j_3}} \hbox{tr} {\bf C}_{\bf
j}^{\bf ll'}= \frac{1}{2^{j_1+j_2+j_3}}\sum_{l_1=0}^{2^{j_1}-1}
\sum_{l_2=0}^{2^{j_2}-1}\sum_{l_3=0}^{2^{j_3}-1}
  [\tilde\epsilon_{\bf j}^{\bf l}]^2,
\end{equation}
and
%eq8
\begin{equation}
N_{\bf j} = \frac{1}{2^{j_1+j_2+j_3}}\sum_{l_1=0}^{2^{j_1}-1}
\sum_{l_2=0}^{2^{j_2}-1}\sum_{l_3=0}^{2^{j_3}-1}
\int\frac{\psi_{\bf j,l}^2({\bf x})}{\bar{n}_g({\bf x})}
 d{\bf x}.
\end{equation}
The physical meaning of eq.(6) is clear. The term
$I_{\bf j}^2$ is the mean power of ${\bf j}$ modes measured from
the observed realization $n_g({\bf x})$, and the term $N_{\bf j}$
is the power on ${\bf j}$ modes due to the Poisson noise. For a
volume-limited survey, the mean galaxy density $\bar{n}_g$ is
independent of the redshift. The Poisson noise power is thus simply
$1/\bar{n}_g$. $P_{\bf j}$ is usually referred to as the DWT power
spectrum.

Paper I showed that for 1-D sample in space $L$ the DWT power
spectrum $P_j$ is a band averaged Fourier spectrum given by
%eq9
\begin{equation}
P_j = \frac{1}{2^j} \sum_{n = - \infty}^{\infty}
 |\hat{\psi}(n/2^j)|^2 P(n),
\end{equation}
where $P(n)$ is the Fourier power spectrum with wavenumber $k=2\pi
n/L$ (for simplicity, we refer to $n$ as the wavenumber below), and
$\hat{\psi}(n)$ is the Fourier transform of the basic wavelet
$\psi(\eta)$
%eq10
\begin{equation}
\hat{\psi}(n)= \int_{0}^{L} \psi(\eta) e^{-i2\pi n\eta}d\eta.
\end{equation}
which is orthogonal to the monopole, i.e. $\hat{\psi}(0)=0$ and
localized in wavenumber space. For the Daubechies 4 wavelet
(Daubechies 1992),
$|\hat{\psi}(n)|^2$ has symmetrically distributed peaks with
respect to $n=0$. The first highest peaks are non-zero in two
narrow ranges centered at $n=\pm n_p$ with width $\Delta n_p$ (see
Fig. 1). Besides the first peak, there are ``side lobes" in
$|\hat{\psi}(n)|^2$ (see also Fig. 1). However, for the Daubechies
4 the area under
the ``side lobes" is not more than 2\% of the first peak.
Therefore $P_j$ is good estimation of the band-averaged Fourier power
spectrum centered at wavenumber
%eq11
\begin{equation}
n_j = n_p 2^j,
\end{equation}
with the band width as
%eq12
\begin{equation}
\Delta n =2^j \Delta n_p.
\end{equation}
For 1-D sample of Ly$\alpha$ forests, the DWT power spectrum is
found to be smoothly related to the Fourier power spectrum by
eqs.(11) and (12) (Feng \& Fang 2000.)

\subsection{Diagonal and off-diagonal modes}

It is straightforward to generalize eq.(9) to 3-D. We have
%eq13
\begin{equation}
P_{\bf j} = \frac{1}{2^{j_1+j_2+j_3}}
  \sum_{n_1 = - \infty}^{\infty}
  \sum_{n_2 = - \infty}^{\infty}
  \sum_{n_3 = - \infty}^{\infty}
  |\hat{\psi}(n_1/2^{j_1})\hat{\psi}(n_2/2^{j_2})
  \hat{\psi}(n_3/2^{j_3})|^2 P(n_1,n_2,n_3).
\end{equation}
Because the cosmic density field is isotropic, the Fourier power
spectrum $P(n_1,n_2,n_3)$ is dependent only on
%eq14
\begin{equation}
n=\sqrt{n_1^2+n_2^2+n_3^2}.
\end{equation}
Obviously, the DWT power spectrum is invariant with respect to the
cyclic permutation of index as
%eq15
\begin{equation}
P_{j_1,j_2,j_3}=P_{j_3,j_1,j_2}=P_{j_2,j_3,j_1}
\end{equation}

Considering eq.(11) and (14), we can formally define a band center
wavenumber $n_j$ corresponding to the 3-D mode ${\bf j}$ as
%eq16
\begin{equation}
n_{\bf j} = n_p \sqrt{(2^{j_1})^2+(2^{j_2})^2+(2^{j_3})^2}.
\end{equation}
For an isotropic random field, the Fourier modes with the same $n$
[eq.(14)] are statistically equivalent. However, the DWT modes
with the same $n_j$ [eq.(16)] are not statistically equivalent,
because the DWT modes are not rotationally invariant. A Fourier
mode $e^{-i(2\pi/L)(n_1x_1+n_2x_2 +n_3x_3)}$ can be obtained by a
rotation of mode $e^{-i(2\pi/L)(n'_1x_1+ n'_2x_2 +n'_3x_3)}$ as
long as $n'^2_1+n'^2_2+n'^2_3 =n_1^2 + n_2^2 + n_3^2$. However,
the DWT modes don't have the same property. Generally, one
cannot transform a mode $(j_1, j_2, j_3)$ to $(j'_1, j'_2, j'_3)$
by a rotation, even when $n_j \simeq n_{j'}$. Because of different
configurations [eq.(2)] between them, the condition $n_j = n_{j'}$
generally does not imply
%eq17
\begin{equation}
P_{\bf j}=P_{\bf j'}.
\end{equation}
This invariance holds only when $(j_1, j_2, j_3)$ is a cyclic
permutation of $(j'_1, j'_2, j'_3)$.

With this property, one can define two types of the DWT power
spectra: 1. the diagonal power spectrum given by $P_{\bf j}$ on
diagonal modes $j_1=j_2=j_3=j$, and 2. off-diagonal power spectrum
given by other modes.

 From eq.(13), the diagonal power spectrum $P_j \equiv P_{j,j,j}$
is related to the Fourier power spectrum by
%eq18
\begin{equation}
P_j =
  \sum_{n_1 = - \infty}^{\infty}
  \sum_{n_2 = - \infty}^{\infty}
  \sum_{n_3 = - \infty}^{\infty}W_j(n_1,n_2,n_3)
   P(n_1,n_2,n_3),
\end{equation}
where the window function $W_j$ is
%eq19
\begin{equation}
W_j(n_1,n_2,n_3)=
\frac{1}{2^{3j}}|\hat{\psi}(n_1/2^{j})\hat{\psi}(n_2/2^{j})
  \hat{\psi}(n_3/2^{j})|^2.
\end{equation}
with the normalization
%eq20
\begin{equation}
\int_{-\infty}^{\infty} W_j(n_1,n_2,n_3) dn_1dn_2dn_3=1
\end{equation}
It should be emphasized that the window function $W_j$ does not
contain any free parameters, and is completely determined by the
basis used for the DWT decomposition. Therefore, it is different
from the ordinary convolution with filter containing adjustable
parameters.

Eq.(11) shows that in the scale space the window function $W_j$ is
localized around $n_1=n_2=n_3= n_p2^j$. Therefore, the diagonal
power spectrum $P_j$ is a band-average of the isotropic Fourier
power spectrum $P(n)$ with the central frequency
$n=\sqrt{3}n_g2^j$. As an example, Fig. 1 demonstrates the linear
diagonal DWT power spectra in three typical cosmological models,
i.e. the $\Lambda$ cold dark matter, ($\Lambda$CDM), $\tau$CDM,
and the standard CDM (SCDM). They are calculated by eq.(18), in
which the $P(n)$ are taken to be the correspondingly linear
isotropic Fourier power spectra (Bardeen et al. 1986). The wavelet
filter $|\hat{\psi}(n)|^2$ [eq.(10)] is displayed in a small frame
box.

For off-diagonal modes, one can also calculate the linear
non-diagonal DWT power spectrum $P_{j_1,j_2,j_3}$ via eq.(13.)
However, in this case, $P_{j_1,j_2,j_3}$ cannot simply be
identified as a band average of the isotropic Fourier power
spectrum $P(n)$ centered at $n=n_j$.

Nevertheless, $n_j$ is useful to calibrate the physical scale of a
given ${\bf j}$. Since we use the DWT decomposition, it is
convenient to label the spatial scale in figures by $j$. However,
the spatial scale for a given ${\bf j}$ actually is dependent on
the size of the sample $L$, i.e. $L/2^j$. Therefore, for samples
with different $L$, the same $j$ do not have the same physical
scale. To avoid the possible confusion caused by this, we label
the figures on the top with the wavenumber $k=2\pi n_j/L$.

\section{Samples}

\subsection{The LCRS galaxies}

We use the Las Campanas redshift survey (Shectman et al. 1996)  to
demonstrate the DWT power spectrum estimator, including both
diagonal and off-diagonal modes. The LCRS consists of 26418
redshifts of galaxies, covering over 720 square degree in six
$1.5^{\circ}\times 80^{\circ}$ strips. Three each  of the strips
are in the north and south galactic caps, respectively. The
galaxies in the LCRS are selected from a CCD catalog obtained in
R-band. The luminosity function has been shown to be best-fitted
by a Schechter function with $M^{*}=-20.29\pm 0.02+5\log{\rm  h}$,
$\alpha=-0.70\pm 0.05$ and $\phi^{*}=0.019\pm0.001$
h$^3$Mpc$^{-3}$ for the absolute magnitude of
$-23.0\leq M-5\log {\rm h} \leq -17.5$(Lin et al. 1996.)

We first convert the heliocentric redshifts of galaxies to comoving
distances $r$ in a Einstein-de Sitter universe.  As the survey depth is
not very large, the differences among the distances given different
models are less than 2\%. This correction has been taken into
account in our calculation.
 Analysis is carried out for the sample with upper cut-off of the
recession velocity 51000kms$^{-1}$, which is approximately the
survey depth $\simeq$ 450 h$^{-1}$ Mpc.
To perform DWT power spectrum analysis, we further enclose each
strip into a cubic box with size of 500 h$^{-1}$ Mpc. Accordingly,
the spatial size resolved by one-dimensional mode on a scale $j$ is
$ 500/2^j$ h$^{-1}$ Mpc.

A slice-like geometry of the strips fills only a very small
fraction of the cubic volume. We define a filling factor by
%eq21
\begin{equation}
f_s=\frac{V_s}{L_{box}^3}
\end{equation}
where $V_s$ is the volume occupied by the survey sample, and
$L_{box}$ is the length of the box. For a given scale ${\bf
j}=(j_1, j_2, j_3)$, there are totally $N_{\bf j}=2^{j_1+j_2+j_3}$
DWT modes. Since the wavelet $\psi_{j,l}$ is
localized in physical space, the WFCs of modes located in the
empty regions will be zero. Consequently, for the scale ${\bf
j}=(j_1, j_2, j_3)$, only $f_sN_{\bf j}$ modes contribute to the
DWT power spectrum. In this case, the summations overing $l_1$,
$l_2$ and $l_3$ in eqs.(7) and (8) run only for the $f_sN_{\bf j}$
modes, and the normalization factor $1/2^{j_1+j_2+j_3}$ is replaced
by $f_sN_{\bf j}$.

The effect of the boundary can be estimated with the so-called
``influence" cone (Pando \& Fang 1998.) That is, if the base
$\psi_{j,l}(x)$ is well localized in the interval $\Delta x$, then
the WFCs $\tilde{\epsilon}_{jl}$ corresponding to position $x_0$
will only measure fluctuations in the interval $[(x_0-\Delta
x)/2^j+1, (x_0+\Delta x)/2^{j+1}]$. If $x_0$ is at the boundary,
the WFCs $\tilde{\epsilon}_{jl}$ measures the difference between
the density contrasts inside and outside the boundary. Thus, for
a zero padding of the empty regions, the power on the boundary
mode would be statistically overestimated. Hence, the uncertainty
(most likely overestimation) of the power spectrum caused by the
boundary modes would be of the order of $N^b_{\bf j}/f_sN_{\bf
j}$, where $N^b_{\bf j}$ is the number of boundary modes on scale
${\bf j}$. One can also pad the empty regions with galaxies
created by a Poisson sampling according to the same radial
selection function as the LCRS. In this case, the difference
between the density contrasts inside and outside the boundary is
erased, so contribute little to the DWT power spectrum and will
cause small underestimation. For small scales $N^b_{\bf j} \ll
f_sN_{\bf j}$, the boundary effect is negligible for both zero
padding and Poisson sampling.

\subsection{N-body simulation and mock LCRS catalogues}

To access the accuracy of recovering power spectrum with the DWT,
we construct mock LCRS samples. First we produce ten realizations
of the N-body simulation for each model of the standard cold dark
matter model
(SCDM), low density flat model ($\Lambda$CDM), and a variant of
SCDM model ($\tau$CDM). The parameters
$(\Omega_0,\Lambda,\Gamma,\sigma_8)$ are taken to be
(1.0,0.0,0.5,0.55) for SCDM, (0.3,0.7,0.21,0.85) for
$\Lambda$CDM, and (1.0,0.0,0.25,0.55) for $\tau$CDM. The
linear power spectrum is taken from the fitting formula given in
Bardeen et al. (1986). We use modified AP$^3$M code (Couchman,
1991) to evolve $128^3$ cold dark matter particles in a periodic
cube of length 256h$^{-1}$Mpc. For the SCDM model, the
starting redshift is taken to be $z_i=15$, and for LCDM and
$\tau$CDM, $z_i=25$, respectively. The force softening parameter
$\eta $ in the
comoving system decreases with time as $\eta \propto 1/a(t)$. Its
initial value is taken to be $\eta=384$ h$^{-1}$\, kpc, and the
minimum value to be $\eta_{\min}=128$ h$^{-1}$\, kpc, corresponding
to 15\% and 5\%, respectively,  of the grid size.  For the
single-step integration of time evolution, we use the
``leap-frog'' scheme, the total number of integration steps down
to $z=0$ is 600.

We first assume an unbiased galaxy distribution relative to
underlying dark matter. The mock catalogues are then constructed
proceeding in the following steps. 1. We locate the observer
randomly in the simulation box.  2. Since the survey depth and
width of the LCRS are larger than our simulations, we replicate
the slice sample periodically be extending to redshift $z \sim
0.2$. 3. From the space coordinate of the galaxy, work out its
redshift and galactic coordinates, for which the angular mask of
the LCRS is further applied. 4. According to R-band luminosity
function and the average sampling rates respectively for 50-fiber
and 112-fiber subset of the observation data, we determine whether
the galaxy is admitted into the mock sample using Poisson
sampling. Similar to the estimation of Fourier power spectrum in
the LCRS catalogue (Lin et.al., 1996), the uneven sampling rate
has little effect on measurements of the DWT Power spectrum.
Finally in step 5, the photometric catalog is then generated by
assigning each galaxy a specific absolute magnitude.

\section{Estimating uncertainties with the DWT power spectrum}

\subsection{Diagonal power spectrum}

In order to demonstrate the reliability of the DWT power spectrum
estimator, we first calculate the diagonal DWT power spectrum for
the simulation ensembles in the SCDM, $\Lambda$CDM, and $\tau$CDM
models. The sample is in a $256^3$ h$^{-3}$ Mpc$^3$ box. The results
are displayed in Fig. 2. For comparison, Fig. 2
also plots the non-linear DWT power spectrum calculated by Eq.(18),
in which the Fourier power spectrum is taken from the Peacock \&
Dodds's fitting formulae (1996.) Obviously, the DWT power spectrum
estimator is excellent to recover the analytic models over a wide
wavenumber range resolved by our simulations, i.e. $0.1 < k < 3$
h Mpc$^{-1}$.

For the magnitude-limited mock samples of the LCRS in real space
with size $500^3$ h$^{-3}$ Mpc$^3$, the estimated diagonal DWT
power spectrum are shown by solid squares in Fig. 3. The error
bars are given by standard 1-$\sigma$ deviations taken over 10
realizations. Comparing with the power spectrum calculated by
Eq.(18), one finds that the DWT power spectrum estimator can
successfully recover the non-linear spectra of the mass fields on
scales $j>3$, i.e., $k > 0.1$ h Mpc$^{-1}$, while the power is
systematically underestimated for $k < 0.1$ h Mpc$^{-1}$

Recall that the Fourier power spectrum also shows a suppression of
power spectrum on large scales. It is generally believed that the
suppression of the Fourier spectrum is due to the following two
reasons: 1. the uncertainty in determination of mean density
(Peacock \& Nicholson 1991; Tadros \& Efstathiou, 1996); 2. the
finite volume of the survey.

The large scale suppression of the DWT power spectrum have
different reasons. For the DWT power spectrum estimator, the
overall mean density is not needed. The uncertainty of the mean
density doesn't affect the DWT power spectrum. As a numerical
test, we computed the DWT power spectrum in an arbitrary selected
thin slab in the simulation box with the same filling factor as
that of the LCRS. We found that the scatter of the DWT power
spectra from different slabs is much less than the suppression
shown in Fig. 2. Accordingly, the uncertainty of the overall mean
density is {\it not} responsible for the suppression of the DWT
power spectrum on large scales.

The suppression of the DWT power spectrum on large scales is
because the diagonal mode $(j,j,j)$ cannot match with the
slice-like geometry of the survey. For the LCRS, the average
thickness in declination is estimated to be $\sim 15$ h$^{-1}$
Mpc, i.e. only $\sim$ 3\% of the side lengths in the other two
dimensions. Therefore, the perturbations on scales $k < 0.2$ h
Mpc$^{-1}$, or $j \leq 3$, in the dimension of declination are
poorly sampled. This leads to an underestimation of power
spectrum with $j \leq 3$ in the direction of declination (\S
3.1). Based on above discussions, for mock samples with the LCRS
survey geometry, the diagonal power spectrum $j_1=j_2=j_3=j$ are
capable of recovering the non-linear power spectrum on scales
$j\geq 3$, but suffers an underestimation on scales $j \leq 3$.
This geometric effect could also be fixed by the off-diagonal
power spectrum as demonstrated in next subsection.

\subsection{Slice-like geometry and off-diagonal power spectrum}

Assuming that $j_1$ points along the direction of declination, $j_2$
the R.A. and $j_3$ the radial or redshift direction, the modes
$(j_1,j_2,j_3)$ which can fit with the LCRS' slice-like geometry
are $2 \leq j_2, j_3 \leq 7$ and $4 \leq j_1 \leq 7$. We use
the off-diagonal power spectrum $P_{j_1,j,j}$ with a fixed
$j_1\geq 4$.

Figure 4 illustrates the off-diagonal DWT spectrum $P_{5,j,j}$
measured from subsets of the sample in a $0.1L \times L\times L$
slab for SCDM and $\Lambda$CDM models respectively, in which
redshift distortion and selection effect have not been taken into
account. Clearly, it shows that the off-diagonal DWT power
spectrum $P_{5,j,j}$ estimator is capable of reproducing the
non-linear power spectrum on large scales. This power spectrum
is in good agreement with the theoretical curves on large scales
as $k \simeq 0.5$ h Mpc$^{-1}$.

Figure 5 plots the off-diagonal DWT spectra $P_{4,j,j}$ and
$P_{5,j,j}$ in real space measured for the mock LCRS samples in
the $\Lambda$CDM model. It again shows that the systematical
underestimation of the powers on large scales is significantly
erased. The off-diagonal DWT spectra $P_{5,j,j}$ and $P_{4,j,j}$
can be used for recovering the non-linear power spectrum of the
mass field on scales as large as $k \simeq 0.15$ h Mpc$^{-1}$.

\subsection{Inhomogeneous effect of selection function}

The purpose of introducing selection function $\bar{n}_g(r)$ is to
eliminate the inhomogeneity and anisotropy of mean number density
of galaxies caused by the sampling of the survey. That is, the
distribution with uniformly mean number density of galaxies can be
recovered by weighting a galaxy at radial coordinate $x_3=r$ by
factor $1/\bar{n}_g(r)$.

However, the effects of selection function actually are two-fold.
On the one hand, it is to eliminate the inhomogeneity of the mean
number density of the galaxy distribution. On the other hand, it
causes an inhomogeneity in the correlations of the power spectrum.
The later can be seen from the band-band correlation in the Paper
I, which showed that, if the selection function is slowly varying,
the observed auto-correlation of power $P_{\bf j}$ is
%eq22
\begin{equation}
\langle P_{\bf j}P_{\bf j} \rangle = \langle P_{\bf j}P_{\bf j}
\rangle_I +
 \frac{1}{2^{2(j_1+j_2+j_3)}}
  \sum_{l_1 = - 0}^{2^{j_1}-1}
  \sum_{l_2 = - 0}^{2^{j_2}-1}
  \sum_{l_3 = - 0}^{2^{j_3}-1}
\left [\frac{1}{\overline{\bar{n}^2_g(x_{3})}}
 +\int\frac{\psi^4_{\bf j,l}({\bf x})}{\bar{n}_g^3(x_3)}d{\bf x}
 \right ],
\end{equation}
where $\langle P_{\bf j}P_{\bf j} \rangle_I$ is the power-power
correlation before the Poisson sampling, and therefore, cyclic
invariant with respect to $(j_1,j_2,j_3)$, i.e.
%eq23
\begin{equation}
\langle P_{j_1,j_2,j_3}P_{j_1,j_2,j_3} \rangle_I =\langle
P_{j_2,j_3,j_1}P_{j_2,j_3,j_1} \rangle_I = \langle
P_{j_3,j_1,j_2}P_{j_3,j_1,j_2} \rangle_I .
\end{equation}
The number $\overline{\bar{n}_g(x_{3})}$ in eq.(22) is the mean of
$\bar{n}_g(x_{3})$ in cell $(l_1,l_2,l_3)$. Since $
\bar{n}_g(x_{3})$ depends only on $x_3$, the last term in the
r.h.s. of eq.(22) violates the permutation invariance. That is, the
correlations $\langle P_{j_1,j_2,j_3}P_{j_1,j_2,j_3} \rangle$ are
on longer cyclic invariant. In other words, the probability
distribution function (PDF) of $P_{\bf j}$ is anisotropic.

To estimate this effect, we calculate the off-diagonal power
spectra $P_{4j4}$ and $P_{44j}$ vs. $j$ for $\Lambda$CDM
simulation sample in real space, and the selection function is
taken to be that for the LCRS galaxies, but  in plane parallel
approximation. The result is shown in Fig. 6. The two spectra
$P_{4j4}$ and $P_{44j}$ show some difference, which is not significant
on all scales $2 \leq j \leq 7$ within 1-$\sigma$. Therefore, with
the DWT estimator, the effect of the anisotropy given by the LCRS
selection function can be ignored at least for the power
measurement on scales similar to $P_{4j4}$ and $P_{44j}$.

\subsection{Redshift distortion and random velocity field of galaxies}

Usually, the redshift distortion is considered to have two effects
on the power spectrum measurement: 1. the enhancing of power on
large scales due to the linear effect of redshift distortion; 2.
the suppressing of power on small scales due to the random motions of
galaxies inside virialized groups and clusters of galaxies. The
general theory of the two effects on the DWT power spectrum is
given in Appendix. An interesting result is that the effect of the
galaxy random velocity field ${\bf v}({\bf x})$ upon the power on
mode ${\bf j}$ is mainly given by the power of the pairwise velocity
dispersion of the same mode. Therefore, the DWT power spectrum can
be employed for
extracting information of the scale dependence of galaxy
velocity fluctuations.

Recall that in the recovery of the real space Fourier power
spectrum from redshift distorted measurement, it is usually
assumed that the peculiar velocity dispersion is
scale-independent. However, theory and measurement of galaxy
pairwise velocity dispersion do not indicate a scale-independent, but
weakly scale-dependent velocity field. Therefore, it is
necessary to estimate the uncertainty caused by the assumption of
the scale-independence of the velocity fluctuations.

Let's use the plane parallel approximation. In this case, the redshift
distortion is determined by the $x_3$-component of the galaxy
velocity field, i.e. $v_3({\bf x})$. Subjecting $v_3({\bf x})$ to a
DWT decomposition, we have
%eq24
\begin{equation}
\tilde{\epsilon}^{V}_{\bf j,l}=
 \int v_3({\bf x})\psi_{\bf j,l}({\bf x})d{\bf x}.
\end{equation}
The WFCs of the velocity field, $\tilde{\epsilon}^{V}_{\bf j,l}$,
describes the radial velocity difference with spatial separation
on scale ${\bf j}$ at position ${\bf l}$. Therefore, it is the
pairwise velocity on scale ${\bf j}$ at position ${\bf l}$.
The redshift distorted DWT power spectrum $P^S_{\bf j}$ is
approximately given by (see Appendix \S A.2)
%eq25
\begin{equation}
P^S_{\bf j} \simeq \frac{1}{1 + S'_{\bf j}P^V_{\bf j}/H^2_0}P_{\bf j},
\end{equation}
where $S'_{\bf j}$ is determined by the geometry of the mode ${\bf
j}$
%eq26
\begin{equation}
S'_{\bf j} =
  \int \psi^2_{\bf j,l}
\left (\frac{\partial\psi_{\bf j,l}}{\partial x_3}\right)^2 d{\bf
x},
\end{equation}
and $P^V_{\bf j}$ is the power of the pairwise velocity dispersion
on scale ${\bf j}$, i.e.
%eq27
\begin{equation}
P^V_{\bf j}=\langle|\tilde{\epsilon}^{V}_{\bf j,l}|^2\rangle_v.
\end{equation}
Therefore, comparing $P^S_{\bf j}$ with $P_{\bf j}$,
we are able to determine $P^V_{\bf j}$.

As an example, we study the assumption of the scale-independence
of galaxy velocity fluctuations. One can show that, if the power
of pairwise velocity is scale-independent, we have
%eq28
\begin{equation}
S'_{ j_1,j_2,j_3}P^V_{j_1,j_2,j_3} \simeq {\rm const},
\hspace{6mm} {\rm for \ a \ fixed} \ j_3
\end{equation}
Therefore, the scale-dependence of the velocity fluctuations can
be detected by off-diagonal DWT power spectrum with a fixed $j_3$
(redshift direction). Figure 7 plots $P_{4,j,4}$ vs. $j$ in
redshift space for the $\Lambda$CDM simulation samples. It can be
seen clearly from Fig. 7 that the DWT power spectrum $P_{4,j,4}$
in redshift space is almost parallel to the theoretical non-linear
spectrum in real space. The difference between the redshift distortion
suppressions of mode $(4,2,4)$ and $(4,7,4)$ is no more than a
factor of 2. That is, the suppressing factor
$1/(1 + S'_{\bf j}) \simeq$ const, and therefore eq.(28)
approximately hold in the scale range of spectrum $P_{4,j,4}$.
Similar results are obtained for models SCDM and $\tau$CDM models.
The assumption of a scale independent power of the pairwise velocity
seems to be reasonable at least for the DWT power spectrum of the
LCRS sample.

Figure 7 also presents the off-diagonal power spectra $P_{4,4,j}$.
Obviously, $P_{4,4,j}$ significantly differs from $P_{4,j,4}$ in
both shape and amplitude of the power. Comparing to
 $P_{4,j,4}$, the amplitude of $P_{4,4,j}$
is enhanced on large scales and suppressed on small scales.

The difference between the spectra $P_{4,j,4}$ and $P_{4,4,j}$ in
redshift space is worth while to look at. For an isotropic
velocity field, $P^V_{j_1,j_2,j_3}$ is cyclically invariant with
respect to the index $(j_1,j_2,j_3)$, i.e.,
%eq29
\begin{equation}
P^V_{4,4,j}= P^V_{4,j,4}.
\end{equation}
The difference between $P_{4,4,j}$ and $P_{4,j,4}$ is simply given
by the geometric factors $S'_{\bf j}$, which are not cyclically
invariant
%eq30
\begin{eqnarray}
S'_{4,4,j} & > & S'_{4,j,4} \hspace{1cm} j>4; \\ \nonumber
S'_{4,4,j} & < & S'_{4,j,4} \hspace{1cm} j<4
\end{eqnarray}
Therefore, using the difference between $P_{4,4,j}$ and
$P_{4,j,4}$ in redshift space we are able to directly measure the
power spectrum of galaxy velocity field, $P^V_{4,4,j}$ or
$P^V_{4,j,4}$.

\section{The DWT power spectrum of LCRS galaxies}

\subsection{The diagonal DWT power spectrum}

With the preparations given in last sections, we can measure the
DWT power spectrum of the LCRS galaxies. As discussed in \S 4.1,
the diagonal power spectrum could provide a robust measurement of
the non-linear power spectrum on scales of $k > 0.1$ h Mpc$^{-1}$.
In Fig. 8, we present the results for the LCRS, where the diagonal
DWT power spectra measured in the six slices are plotted by
scatter symbols as indicated in the figure. The average value over
the six strips, and over the three north as well as three south strips
are also displayed by the connected lines, respectively.

Although the geometry of the six strips are actually quite
different, and the filling factor is only about 0.9\%, Fig. 8
indicates that the scatter of the power on scales of $0.1 < k <
2$ h Mpc$^{-1}$ is rather small. Comparing with Fig. 3, we find
that the scatter of the power in Fig. 8 is within the range of
1-$\sigma$ deviations given by the mock LCRS samples. Therefore,
most of the scatters in Fig. 8 are probably from the cosmic
variance. This result implies that the DWT power spectrum
estimator is insensitive to the geometry of survey and filling
factor. This feature  allows us to directly compare the DWT power
spectrum measurements from different surveys, or different region
in a survey.

As indicated in Fig. 8, the DWT power spectrum is capable of probing
the fluctuation power on scales as small as $k= 2$ h Mpc$^{-1}$
($j=7$). This is partially because the DWT decomposition is free
from the ``aliasing effect" which may lead to underestimation of
power on the scale of grid size (Jing 1992, Baugh \& Efstathiou,
1994).  At the scale $j=6$, the number of effective modes is
$f_sN_j \simeq 2500$
comparable to galaxy number in each slice, and thus the error
bars for $j\geq 6$ are mainly from the Poisson noise. Even on the
smallest scale $j=7$ the Poisson correction is still much smaller
than the detected power. Therefore, the diagonal DWT power
spectrum from $k \simeq 0.1$ to 2 h Mpc$^{-1}$ is qualified for
model discrimination.

\subsection{Fitting values of the bias parameter and velocity dispersion}

In order to find the best values of the bias parameter and
velocity dispersion, we compare the detected LCRS diagonal DWT
power spectrum (Fig. 8) with the fitting spectrum in redshift
space, $P^S_{\bf j}$, given by eq.(13), i.e.
%eq31
\begin{equation}
P^S_{\bf j} = \frac{1}{2^{j_1+j_2+j_3}}
  \sum_{n_1 = - \infty}^{\infty}
  \sum_{n_2 = - \infty}^{\infty}
  \sum_{n_3 = - \infty}^{\infty}
  |\hat{\psi}(n_1/2^{j_1})\hat{\psi}(n_2/2^{j_2})
  \hat{\psi}(n_3/2^{j_3})|^2 P^S(n_1,n_2,n_3),
\end{equation}
where $P^S(n_1,n_2,n_3)$ is the Fourier power spectrum in redshift
space. For typical CDM models, $P^S(n_1,n_2,n_3)$ can be taken
from the fitting formula given by Peacock and Dodds (1994). It is
%eq32
\begin{equation}
P^S(k)=b^2P_m(k)G(\beta, k\sigma_v)
\end{equation}
where $P_m(k)$ is the non-linear mass power spectrum in real
space, $\beta=\Omega^{0.6}/b$ is redshift distortion parameter,
$b$ the scale-independent linear biasing parameter, $\sigma_v$ is
1-D peculiar velocity dispersion. $G(\beta,k\sigma_v)$
describes the suppression of redshift distortion.

Exactly speaking, one cannot simply generalize eq.(13) to eq.(31),
because the redshift distortion of the DWT power spectrum is
different from the Fourier power spectrum. The pairwise velocity
suppression of the DWT modes should be calculated based on
eq.(25), rather than via the Fourier modes. Moreover, eq.(32) is
given by averaging over the directions of Fourier modes, and
therefore, it doesn't match the anisotropic mode $(j_1,j_2,j_3)$
of eq.(31). Nevertheless, the fitting of eq.(31) will provide
worth information for diagonal modes, as the power of the galaxy
random velocity field is approximately scale-independent (\S 4.4).

We treat $b$ and $\sigma_v$ in eq.(31) as free parameters, and fit
the power spectrum eq.(31) to the detected DWT power spectrum of
LCRS. The result is shown in Fig. 9, in which the 1-$\sigma$ error
bars of the observed DWT power spectrum are given by the scatter
over the observed 6 slices themselves. As have been discussed in
\S 4.1, the DWT power spectrum at large scales ($j\le 3$) are
underestimated due to the slice-like geometric effect. We
approximately compensate this effect in the observed DWT power
spectrum by using the fractional damping of amplitude calculated
from the mock samples in redshift space relative to the expected
values from the non-linear fitting formula (Eq.(32)) in each
model respectively. The least square best fitting parameters $b$
and $\sigma_v$ for each of models are presented in Table 1.

\begin{table*}
\begin{center}
\centerline{Table 1}
\bigskip
\begin{tabular}{cccclc}
\hline\hline
  Model& b & $\beta(P_{jjj})$ & $\beta(P_{5jj})$ &
    $\sigma_v$ km s$^{-1}$ $(P_{jjj})$
    & $\sigma_v$ km s$^{-1}$($P_{5jj}$)\\
  \hline
 SCDM      & $1.78\pm 0.30$ & $0.56\pm 0.09$ & 0.50 & $306\pm 106$  & 397 \\
 $\tau$CDM & $1.50\pm 0.18$ & $0.67\pm 0.08$ & 0.59 & $135\pm 42$ & 217  \\
 $\Lambda$CDM & $1.06\pm 0.13$ & $0.46\pm 0.06$ & 0.41 &  $250\pm 72$ &
344 \\\hline
\end{tabular}
%\caption{}
\end{center}
\end{table*}

The SCDM and $\Lambda$CDM parameter $\beta$ given in the Table 1
is about the same as other measurements, such as
$\beta=0.52\pm 0.13$ for the 1.2Jy IRAS survey (Cole et al. 1995);
$\beta=0.47\pm 0.16$ (Tadros et al. 1999) and
$\beta=0.41^{+0.13}_{-0.12}$ (Hamilton et al. 2000)
for IRAS Point Source Catalog Redshift Survey (PSCz). Actually,
our estimated $\beta$ in each CDM model is very close to the
linear value $\sigma_8\Omega_0^{0.6}$. This result
is consistent with the theory of redshift distortion in the DWT
presentation, which shows that the linear effect on the diagonal
DWT power spectrum is about the same as that given by eq.(32)
(see, Appendix \S A.1.)

The 1-D velocity dispersions given in Table 1 look lower than
other measurements in the LCRS galaxies. They are $v_{12}=570\pm
80$kms$^{-1}$ (Jing et al. 1998), and $363\pm 13$ km s$^{-1}$
(Landy, Szaley and Broadhurst 1998). Yet, the first
measurement is based on {\it a priori} infall model of galaxies.
In the second measurement $v_{12}$ is given by the width of an
exponential distribution, or a Lorentzian distribution in $k$
space. On the other hand, $\sigma_v$ in eq.(31) is assumed to be the
variance of a Gaussian distribution of pairwise velocity. To fit
a given power suppression by Lorentzian and Gaussian
distributions, $v_{12}$ will always be larger than $\sigma_v$.
For instance, if the suppression factor is equal to about 3 as
shown by the spectra of Fig. 7. $v_{12}/\sigma_v$ will be $\sim
1.4$. Therefore, the values of $\sigma_v$ given by the fitting of
the SCDM and $\Lambda$CDM basically are consistent with the
second measurement.

\subsection{Off-diagonal DWT power spectrum}

As an example we detected the off-diagonal spectrum $P_{5jj}$ of
the LCRS galaxies. The result is presented in Fig. 10.

As mentioned in last section, eq.(31) cannot be used to fit with
off-diagonal DWT power spectrum, as $P^S(k)$ is isotropic. One can
expect that the fitting parameters of the spectrum eqs.(31) and
(32) with an off-diagonal power spectrum would be systematically
different from those of the diagonal power spectrum. This
systematic difference actually contains valuable information. The best
fitted spectra of $P_{5jj}$ for models SCDM, $\tau$CDM and
$\Lambda$CDM are also shown in Fig. 10. The best-fitting
parameters are given in Table 1, which clearly show the
difference between the parameters of diagonal and off-diagonal
spectrum. Now we analyze the physical meaning of the difference.

From Appendix A.1, the linear redshift distortion of spectrum
$P_{\bf j}$ is determined by the factor $\beta S_{\bf j}$, where
$S_{\bf j}$ is a geometric factor given by eq.(A4). The linear
redshift distortion is significant on large scale. Moreover, the
geometric factor satisfies
%eq33
\begin{equation}
S_{5jj} < S_{jjj} \hspace{1cm} {\rm for} \ \  j <5.
\end{equation}
Therefore the spectrum $P_{5jj}$ is less affected by the linear
redshift distortion than $P_{jjj}$. Thus, if fitting both
$P_{jjj}$ and $P_{5jj}$ with the {\it same} fitting formula eq.(31),
the best-fitting $\beta$ for $P_{5jj}$ will be lower than those of
$P_{jjj}$. Table 1 shows the systematic lowering of $\beta(P_{5jj})$
than $\beta(P_{jjj})$.

The effect of galaxy pairwise velocity on spectrum $P_{\bf j}$ is
determined by the power $P^V_{\bf j}S'_{\bf j}$ (Appendix A.2). If
the power of pairwise velocity is approximately scale-independent,
we have
%eq34
\begin{equation}
P^V_{5jj}S'_{5jj} > P^V_{jjj}S'_{jjj}.
\end{equation}
That is, the spectrum $P_{5jj}$ is strongly affected by the random
velocity field than $P_{jjj}$. Thus, if fitting both $P_{jjj}$ and
$P_{5jj}$ with the same fitting formula eq.(31), the best-fitting
$\sigma_v$ for $P_{5jj}$ will be larger than those of $P_{jjj}$.
This is shown in Table 1. The significant difference between the
best-fitting results of $P_{jjj}$ and $P_{5jj}$ shows that the
off-diagonal power spectrum has the capability of measuring the
galaxy velocity field from the observational data.

\section{Conclusions}

The DWT power spectrum estimator is studied with the galaxy
distributions of the Las Campanas redshift survey, and the relevant
mock and simulation samples in models of the SCDM, $\tau$CDM and
$\Lambda$CDM.

The DWT estimator for 3-D samples provide two types of
spectra with respect to diagonal and off-diagonal modes,
respectively. The two types of the DWT local modes have different
configuration and invariance, and therefore, the diagonal and
off-diagonal DWT power spectra are flexible for studying
direction-dependent properties.

Using mock and simulation samples, we studied the effect of the
irregular geometry of the survey, the inhomogeneity caused by the
selection function, the scale-dependence of the redshift
distortion due to the random velocity field of galaxies etc. To
estimate the uncertainty and errors caused by these effects, the
localization of the DWT modes in phase space are essential.
For instance, the off-diagonal power spectra are capable of
measuring the statistical features of
the random velocity field of galaxies on mode ${\bf j}$, because
the DWT power on mode ${\bf j}$ is mainly affected by the power of
the velocity fluctuations on mode ${\bf j}$.

The DWT power spectrum from the six slices of the LCRS galaxies
are very stable over the scale range $0.06 \leq k \leq 2.0 $h
Mpc$^{-1}$, in spite of the fact that the filling factor of the LCRS
galaxies is
only about 0.9\% (in comparison, SDSS's filling factor is about
17\%.) To fit the LCRS DWT power spectrum with CDM models, we find
that the redshift distortion parameter $\beta$ and the 1-D
velocity dispersion $\sigma_v$ are consistent with results of the
Fourier power spectrum. Therefore, the DWT estimator can give a
robust measurement of the banded Fourier power spectrum.

More interesting, the differences between the best-fitting parameters
of LCRS' diagonal and off-diagonal DWT power spectra are found to
be significant. This shows that the redshift space behaviors of the
diagonal and off-diagonal DWT power spectra are different.  With
the diagonal and off-diagonal DWT power spectra, we would be able
to compare the pairwise velocity dispersion on different modes.
Therefore, the difference between diagonal and off-diagonal power
spectra would be valuable to constrain models.

\acknowledgments

LLF and YQC acknowledges support from the National Science
Foundation of China (NSFC) and National Key Basic Research Science
Foundation. We thank anonymous referee for helpful comments.

\appendix

\section{Redshift distortion in the DWT representation}

\subsection{Linear redshift distortion}

If the density field is viewed in redshift space, the observed
radial position is given by the radial velocity consisting of the
uniform Hubble flow and the peculiar motion ${\bf v}({\bf x})$. In
the linear regime of clustering, the effect of redshift distortion
can be described by a linear mapping from density contrast
$\delta({\bf x})$ in real space to redshift space
%eqA1
\begin{equation}
\delta^S({\bf x})={\bf S}\delta({\bf x}),
\end{equation}
where ${\bf S}$ is the linear redshift distortion operator. In
plane-parallel approximation, it is
%eqA2
\begin{equation}
{\bf S}=1+ \beta \frac{\partial^2}{\partial x_3^2}\nabla^{-2},
\end{equation}
where $\beta$ is the redshift distortion parameter, and $x_3$ is
on the direction of redshift.

The differential operator in eq.(A2) is quasi-diagonal when
decomposed into the DWT basis (Farge et al. 1996.) Thus, similar to
eqs.(2) and (3), $\delta^S({\bf x})$ can be decomposed into
%eqA3
\begin{equation}
\tilde{\epsilon}_{\ \ \bf j}^{S, \bf l} \simeq (1+ \beta S_{\bf
j}) \tilde{\epsilon}_{\bf j}^{\bf l},
\end{equation}
where the coefficient $S_{\bf j}$ is given by
%eqA4
\begin{equation}
S_{\bf j}= \int \psi_{\bf j,l}({\bf x}) \frac{\partial^2}{\partial
x_3^2}\nabla^{-2}\psi_{\bf j,l}({\bf x}) d{\bf x}.
\end{equation}
This integral is independent of ${\bf l}$. For diagonal modes
${\bf j}=(j,j,j)$, we have
%eqA5
\begin{equation}
S_{jjj} =1/3.
\end{equation}
Using eqs.(5) and (6), the relation between the DWT power spectra
in redshift space, $P^S_{\bf j}$ and real space $P_{\bf j}$ is
given by
%eqA6
\begin{equation}
P^S_{\bf j}=(1+ \beta S_{\bf j})^2P_{\bf j}.
\end{equation}
For diagonal DWT power spectrum, we have
%eqA7
\begin{equation}
P^S_{jjj}=(1+ \frac{1}{3}\beta )^2P_{jjj}.
\end{equation}
The redshift distortion of the Poisson correction are not considered
in eqs.(6) and (7). The presice theory of the redshift distortion of
the Poisson correction requires to calculate the selection function
by the DWT decomposition. We will not give it here. In a rough
approximation, the effect of the Poisson correction can be
estimated by
%eqA8
\begin{equation}
P^S_{\bf j}=(1+ \beta S_{\bf j})^2I_{\bf j}^2-N_{\bf j}=
   (1+ \beta S_{\bf j})^2(P_{\bf j}+N_{\bf j})-N_{\bf j}.
\end{equation}

\subsection{Effect of random velocity field}

In non-linear regime of clustering, the redshift distortion
generally is estimated by assuming a random velocity field ${\bf
v}({\bf x})$, which describe the velocity dispersion of galaxies.
The average of ${\bf v}({\bf x})$ is zero, $\langle {\bf v}({\bf
x}) \rangle_v=0$, where $\langle ...\rangle_v$ is for the average
over ensemble of velocity fields. The covariance of the density
contrast in redshift space is then given by
%eqA9
\begin{equation}
\langle \langle \delta(x_1, x_2, x_3 + v_3({\bf x})/H_0)
   \delta(x'_1, x'_2, x'_3 + v_3({\bf x'})/H_0)\rangle_v \rangle,
\end{equation}
Thus, the DWT power spectrum in redshift space is
%eqA10
\begin{equation}
P^S_{\bf j}=\langle \langle [\tilde{\epsilon}^{S,{\bf l}}_{\ \ \bf
j}]^2
  \rangle_v\rangle -N_{\bf j},
\end{equation}
where
%eqA11
\begin{equation}
\tilde{\epsilon}^{S,{\bf l}}_{\ \ \bf j}=\int
 \delta(x_1,x_2, x_3 + v_3({\bf x})/H_0)\psi_{\bf j,l}({\bf x})d{\bf x}.
\end{equation}

Using an auxiliary variable $J$, eq.(A10) can be rewritten as
%eqA12
\begin{equation}
 \tilde{\epsilon}^{S,{\bf l}}_{\ \ {\bf j}} =\int d{\bf x} \left.
 \delta \left (x_1, x_2, x_3 - i\frac{\delta}{\delta J}\right )
 e^{iJv_3({\bf x})/H_0}
  \psi_{\bf j,l}({\bf x})  \right |_{J=0}.
\end{equation}
Thus, we have
%A13
\begin{eqnarray}
\lefteqn{ \langle [\tilde{\epsilon}^{S,{\bf l}}_{\ \ \bf j}]^2
\rangle_v = }
   \\ \nonumber
 & &  \int d{\bf x} d{\bf x'}
 \delta \left (x_1, x_2, x_3 - i\frac{\delta}{\delta J}\right )
 \delta \left (x'_1, x'_2, x'_3 - i\frac{\delta}{\delta J'}\right )
   \\ \nonumber
 & & \left. \langle e^{iJv_3({\bf x})/H_0+ iJ'v_3({\bf x'})/H_0}\rangle_v
  \psi_{\bf j,l}({\bf x})\psi_{\bf j,l}({\bf x'}) \right |_{J, J'=0}
\end{eqnarray}

In the lowest order of non-zero correction, we have
%A14
\begin{eqnarray}
\lefteqn {\langle e^{iJv_3({\bf x})/H_0+ J'v_3({\bf
x'})/H_0}\rangle_v \simeq}
   \\ \nonumber
 & & 1- \frac{1}{2H^2_0}[J^2 \langle v_3^2({\bf x})\rangle_v +
   J^{'2} \langle v_3^2({\bf x'})\rangle_v +
   2JJ'\langle v_3({\bf x}) v_3({\bf x'})\rangle_v]
\end{eqnarray}
In this case, the redshift distorted power spectrum is
%eqA15
\begin{equation}
P^S_{\bf j}=\left (1 - \frac{1}{H^2_0}S'_{\bf j}
\langle|\tilde{\epsilon}^{V}_{\bf j,l}|^2\rangle_v \right )P_{\bf j},
\end{equation}
where $\tilde{\epsilon}^{V}_{\bf j,l}$ is the WFCs of velocity
field, i.e.
%eqA16
\begin{equation}
\tilde{\epsilon}^{V}_{\bf j,l}=
 \int v_3({\bf x})\psi_{\bf j,l}({\bf x})d{\bf x}.
\end{equation}
The geometry factor $S'_{\bf j}$ is positive, and is given by
%eqA17
\begin{equation}
S'_{\bf j} =
  \int \psi^2_{\bf j,l}
\left (\frac{\partial\psi_{\bf j,l}}{\partial x_3}\right)^2 d{\bf
x}
\end{equation}
When the correction term is large, we have an approximation as
%eqA18
\begin{equation}
P^S_{\bf j} \simeq \frac{1}{1 + S'_{\bf j}
\langle|\tilde{\epsilon}^{V}_{\bf j,l}|^2\rangle_v/H^2_0 }P_{\bf j},
\end{equation}

If the velocity field is homogeneous and isotropic,
$\langle|\tilde{\epsilon}^{V}_{\bf j,l}|^2\rangle_v$ is ${\bf
l}$-independent. It is the power of velocity fluctuations on scale
${\bf j}$, i.e.
%eqA19
\begin{equation}
P^V_{\bf j}=\langle|\tilde{\epsilon}^{V}_{\bf j,l}|^2\rangle_v.
\end{equation}
Thus, we have
%eqA20
\begin{equation}
P^S_{\bf j} \simeq \frac{1}{1 + S'_{\bf j}P^V_{\bf j}/H^2_0
}P_{\bf j}.
\end{equation}
Therefore, the effect of the galaxy random velocity field upon the
power $P_{\bf j}$ on scales ${\bf j}$ is mainly given by the power of
the $v_3$ fluctuations on the same scale.

Similar to eq.(A8), the redshift distortion of the Poisson correction
can be estimated by
%A21
\begin{equation}
P^S_{\bf j}=\left (1 - \frac{1}{H^2_0}S'_{\bf j}
\langle|\tilde{\epsilon}^{V}_{\bf j,l}|^2\rangle_v \right )
(P_{\bf j}+N_{\bf j})-N_{\bf j}.
\end{equation}

\newpage

%fig1
\begin{figure}
\begin{center}
\vspace{20.0cm} \includegraphics{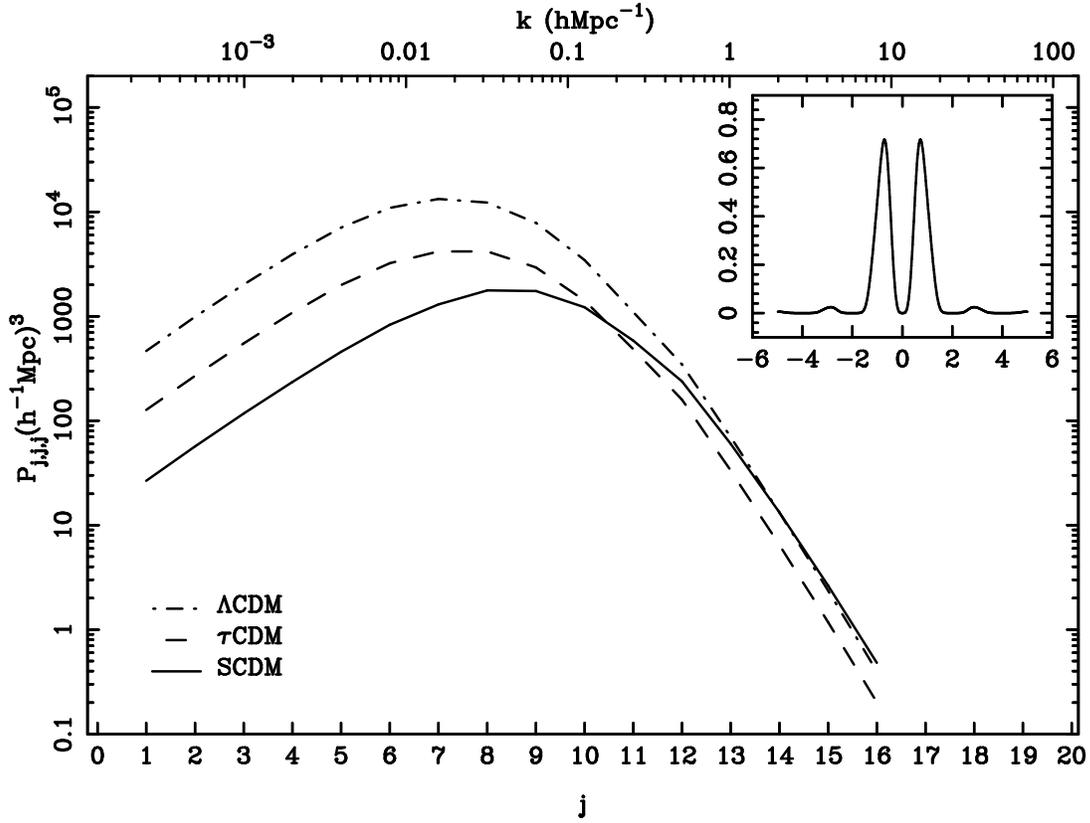} \caption {Linear diagonal DWT
power spectrum in three typical cosmological models:
$\Lambda$CDM, $\tau$CDM and SCDM. The wavelet filter
$|\hat{\psi}(n)|^2$ is displayed in a small frame.} \label{Fig1}
\end{center}
\end{figure}

\clearpage

%fig2
\begin{figure}
\begin{center}
\vspace{20.0cm} \includegraphics{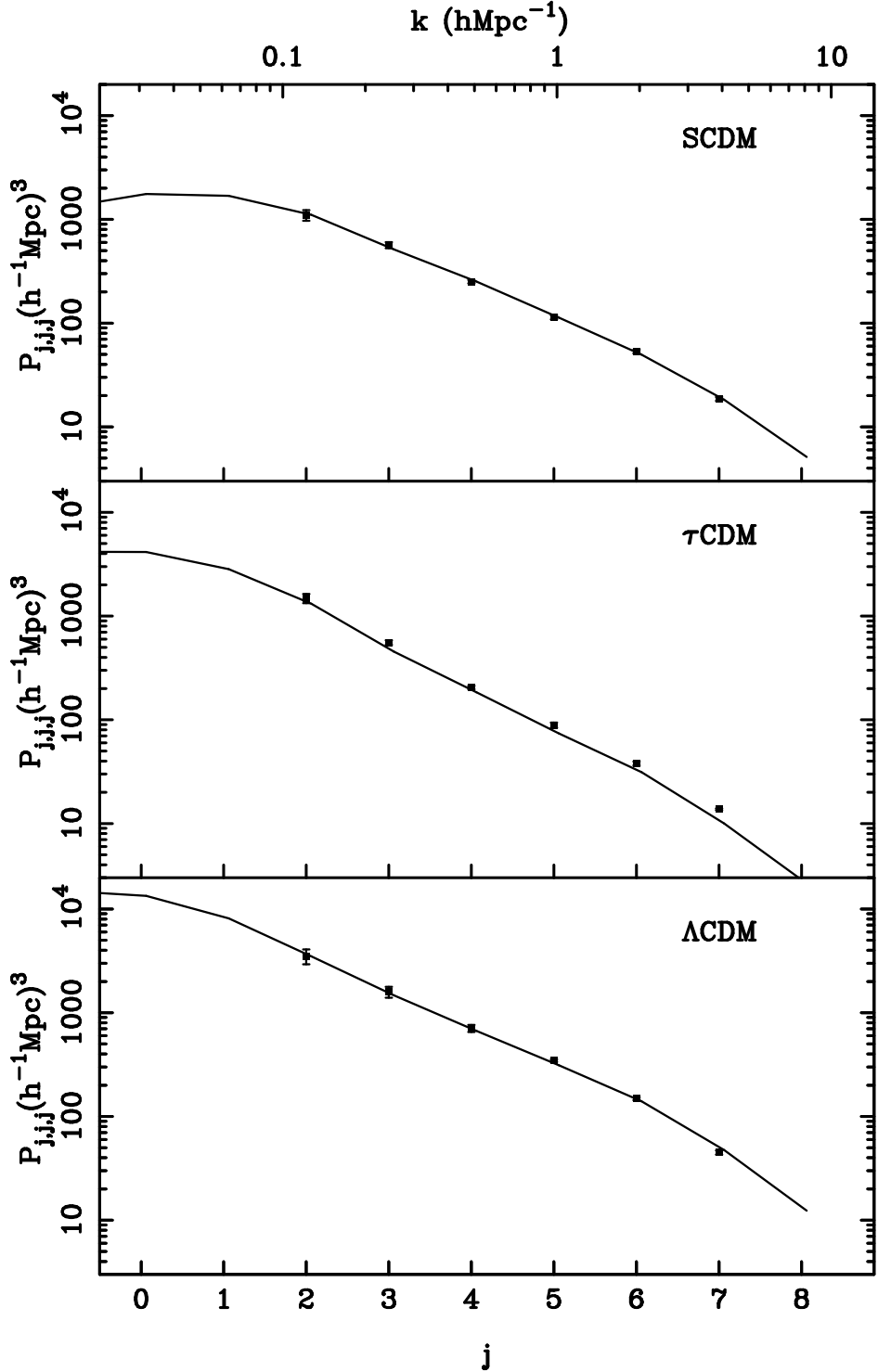} \caption{The non-linear diagonal DWT power
spectrum of simulation samples (solid square) and the
corresponding DWT power spectrum (solid line) calculated by
eq.(18), in which the non-linear Fourier power spectrum
$P(n_1,n_2,n_3)$ are taken from Peacock \& Dodds's fitting
formulae (1994). The error bars are given by 1-$\sigma$
deviations obtained from 10 realizations. } \label{Fig2}
\end{center}
\end{figure}

%fig3
\begin{figure}
\begin{center}
\vspace{20.0cm} \includegraphics{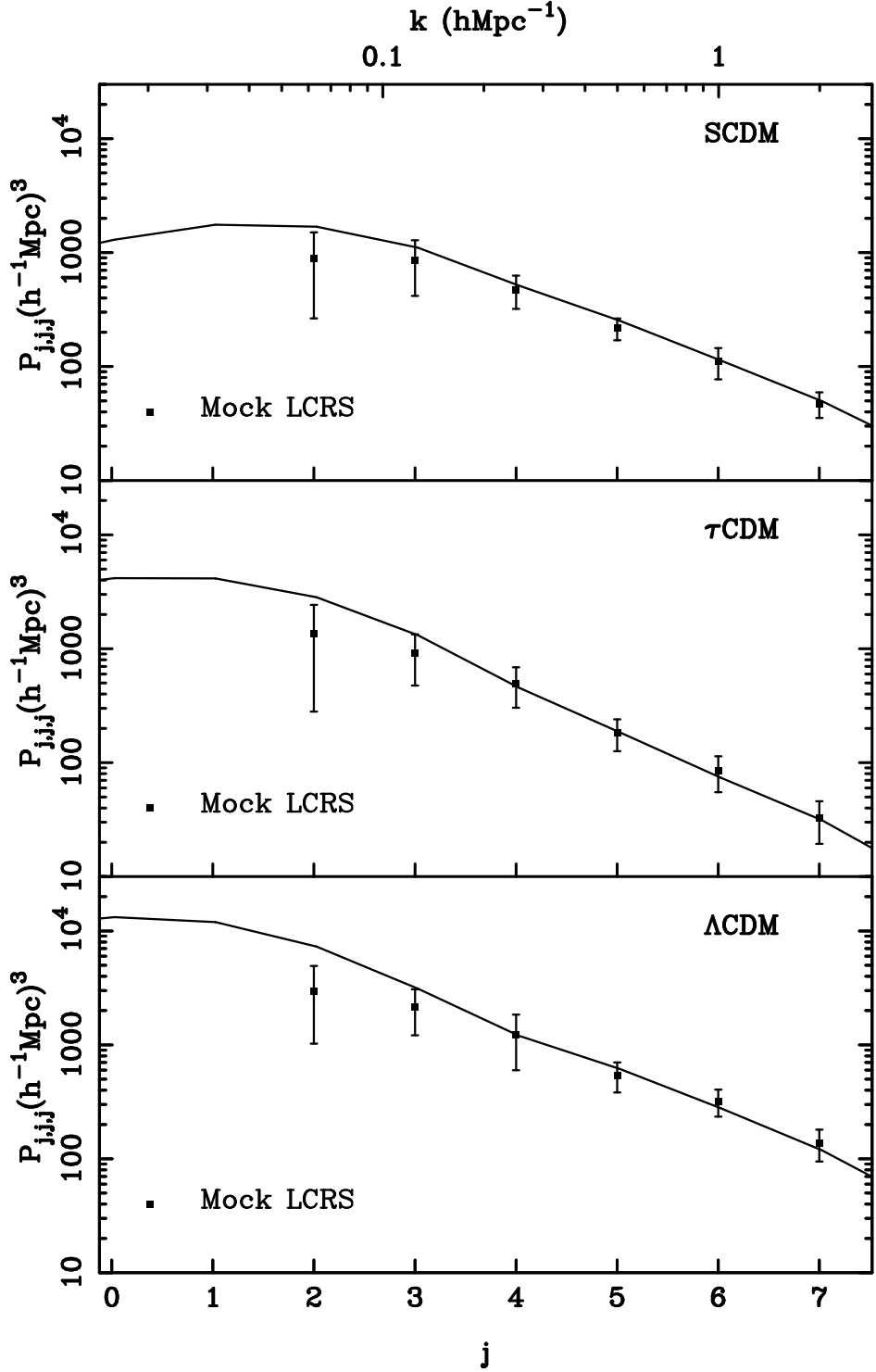} \caption{The diagonal DWT power spectrum of
the mock flux limited LCRS catalogue in real space for models
SCDM (upper panel), $\tau$CDM (central panel) and $\Lambda$CDM
(lower panel). The non-linear DWT power spectra (solid line) are
calculated by eq.(18), in which the Fourier spectrum
$P(n_1,n_2,n_3)$ is taken from the Peacock-Dodds fitting
formulae. The error bars are 1-$\sigma$ variance obtained from 10
realizations for each slice. }\label{Fig3}
\end{center}
\end{figure}

%fig4
\begin{figure}
\begin{center}
\vspace{20.0cm} \includegraphics{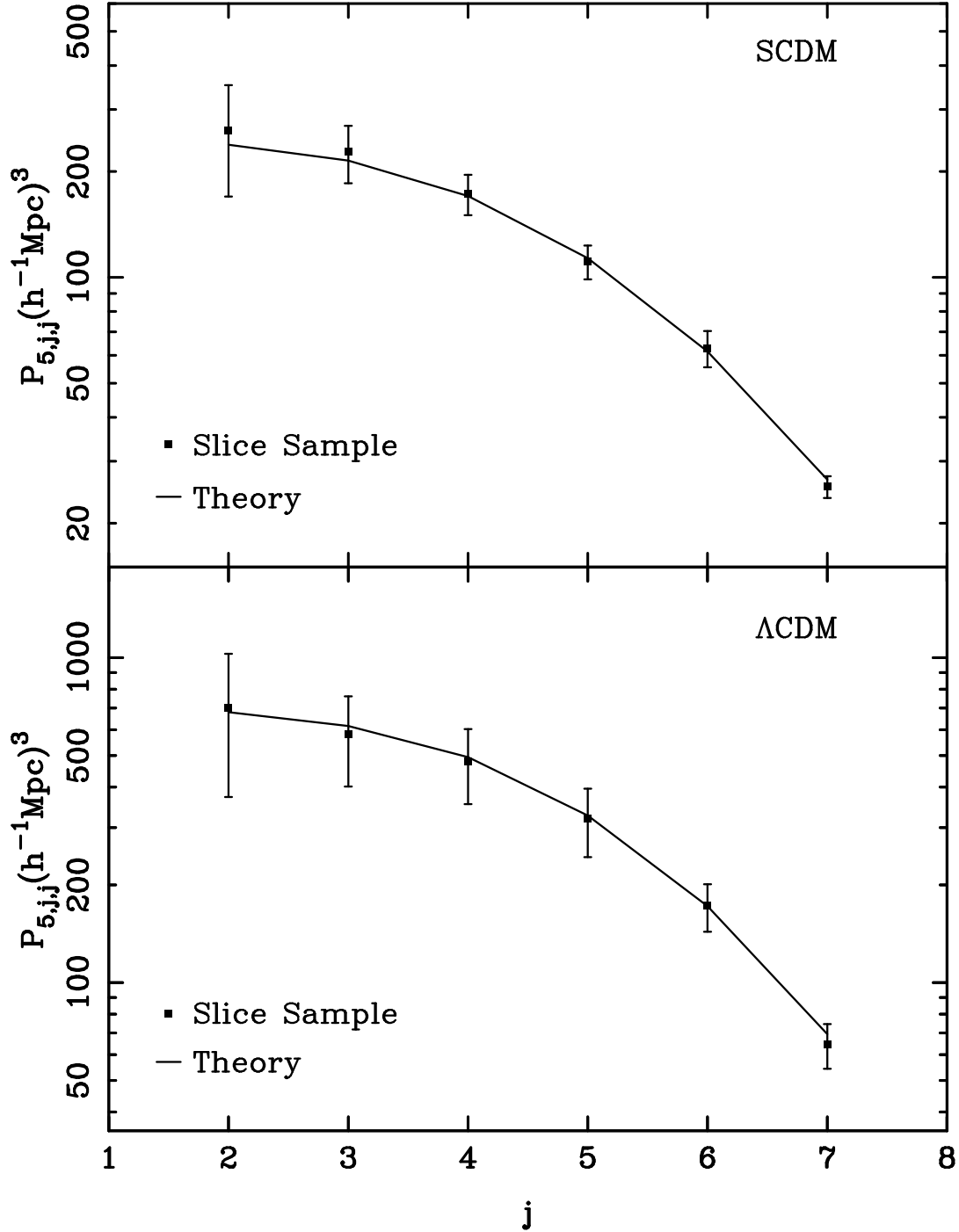} \caption{The off-diagonal DWT spectrum
$P_{5,j,j}$ for SCDM and $\Lambda$CDM simulation slice samples in
real space.  Selection effect is not applied. The dimension in
$x$-axis is thin ($x=0.1L$). The solid line is the spectrum
calculated by eq.(13), in which the Fourier spectrum
$P(n_1,n_2,n_3)$ is taken from the Peacock-Dodds fitting
formulae.} \label{Fig4}
\end{center}
\end{figure}

%fig5
\begin{figure}
\begin{center}
\vspace{20.0cm} \includegraphics{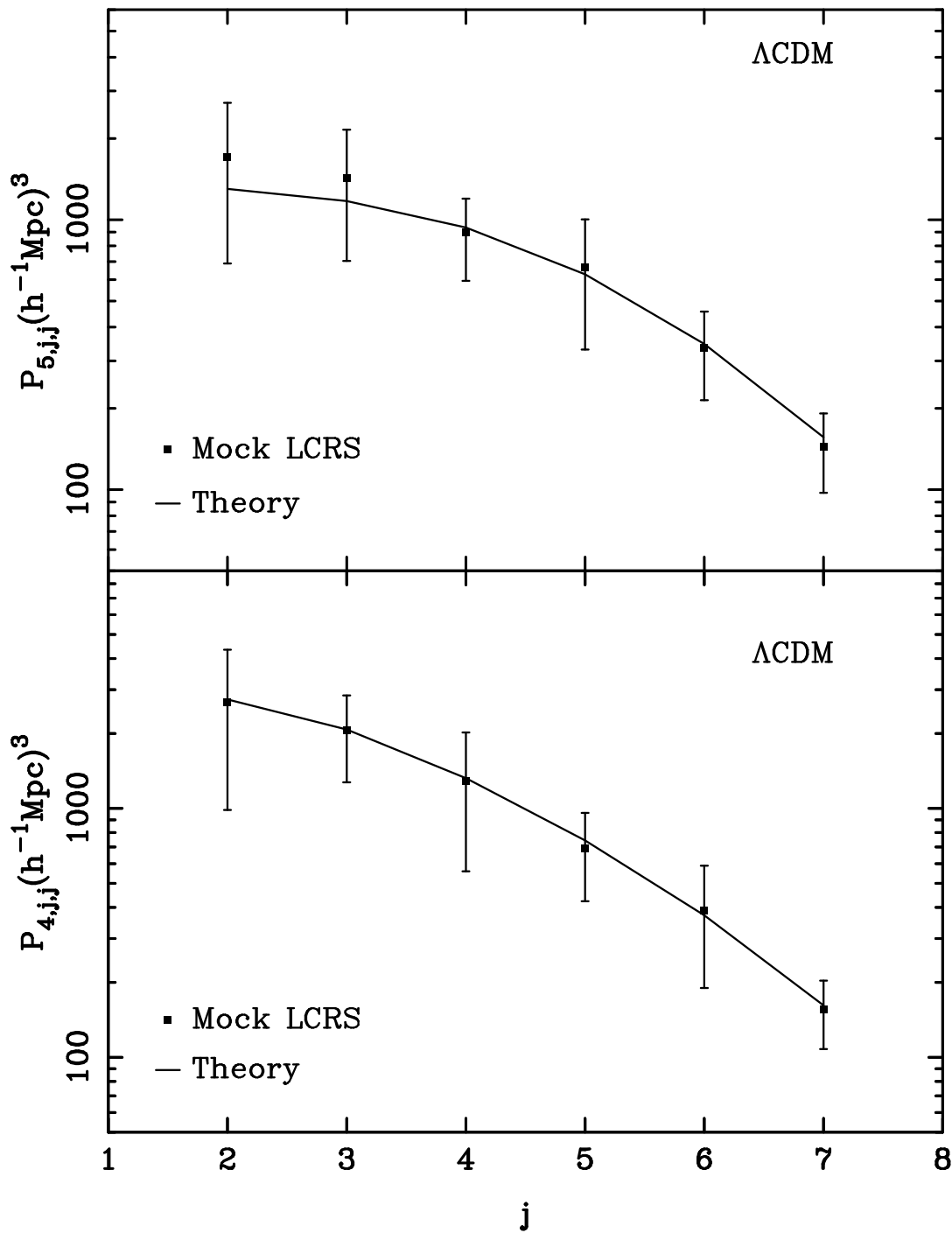} \caption{The off-diagonal DWT spectra
$P_{5,j,j}$ and $P_{4,j,j}$ for $\Lambda$CDM mock LCRS samples in
real space. The solid lines are the spectrum calculated by
eq.(13). } \label{Fig5}
\end{center}
\end{figure}

%fig6
\begin{figure}
\begin{center}
\vspace{20.0cm} \includegraphics{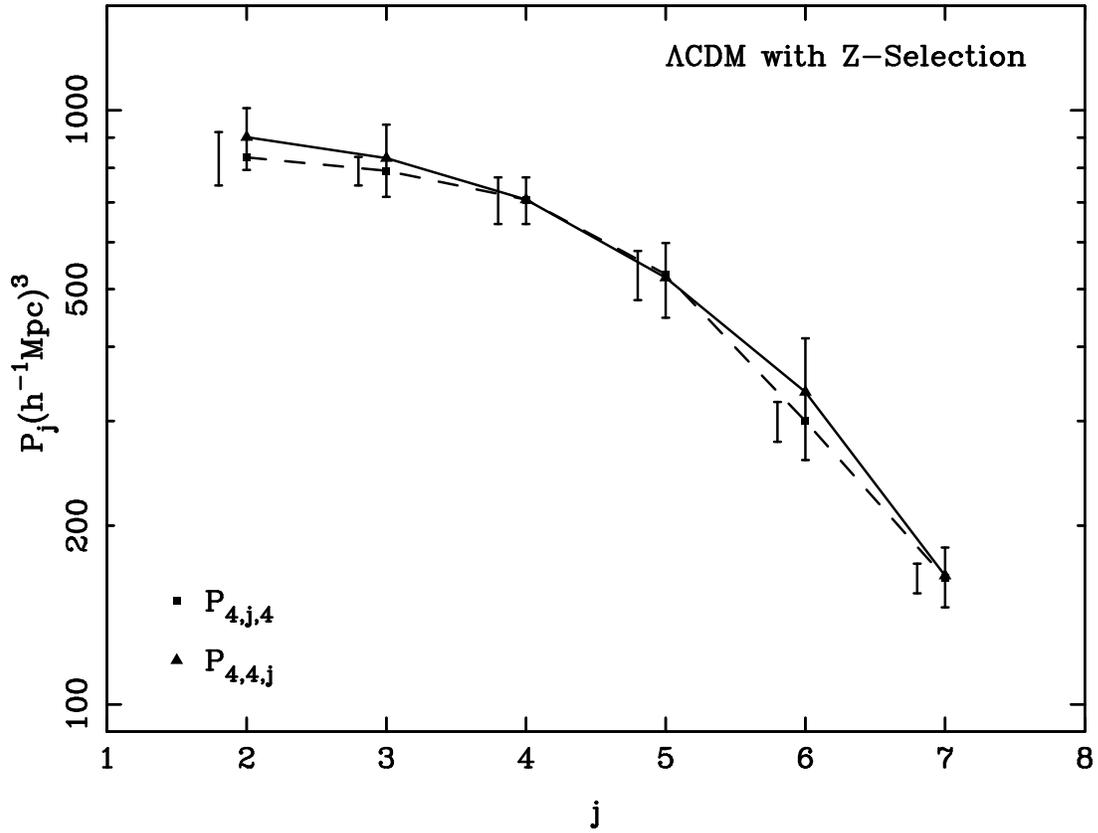}  \caption{The off-diagonal power
spectra $P_{4j4}$ and $P_{44j}$ vs. $j$ for $\Lambda$CDM
simulation sample in real space, but the selection function is
taken account by plane parallel approximation. For clarity, the
error bars of $P_{4j4}$ are shifted left by a factor of 0.2.}
\label{Fig6}
\end{center}
\end{figure}

%fig7
\begin{figure}
\begin{center}
\vspace{20.0cm} \includegraphics{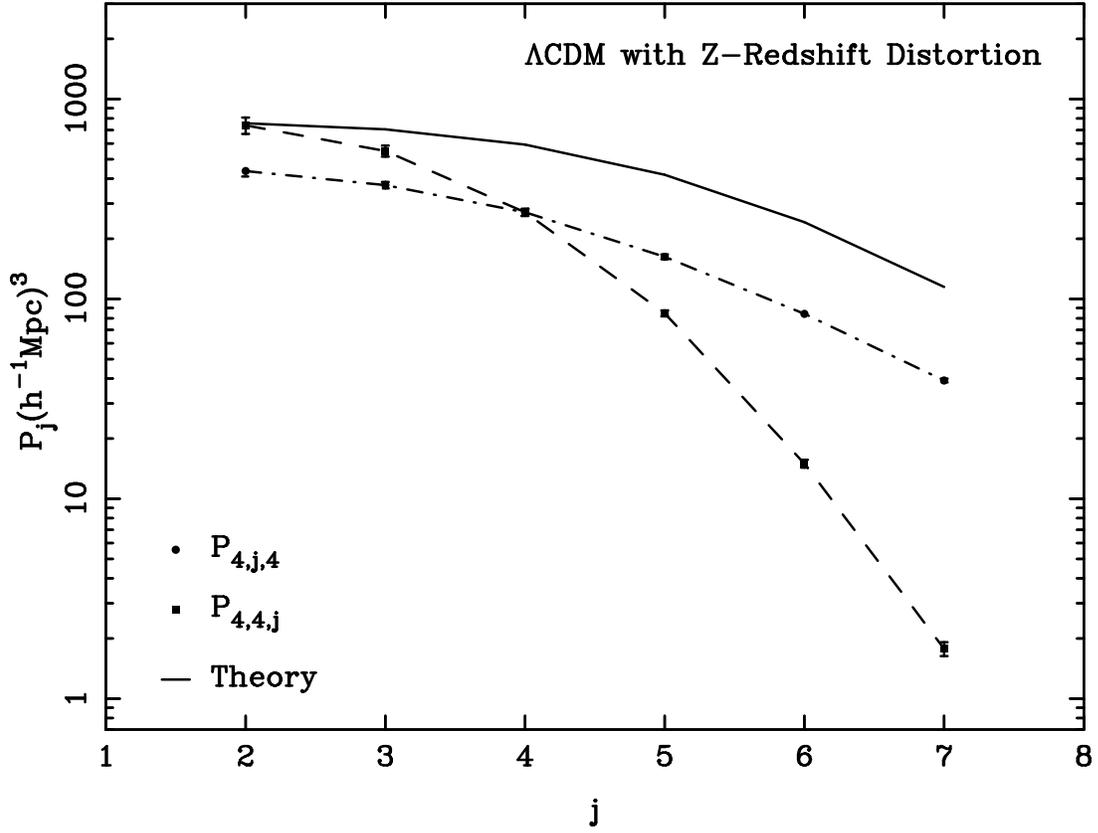} \caption{The off-diagonal power
spectra $P_{4j4}$ and $P_{44j}$ vs. $j$ for $\Lambda$CDM
simulation sample in redshift space. The redshift distortion is
taken account by plane parallel approximation. Selection function
is not applied.} \label{Fig7}
\end{center}
\end{figure}

%fig8
\begin{figure}
\begin{center}
\vspace{20.0cm} \includegraphics{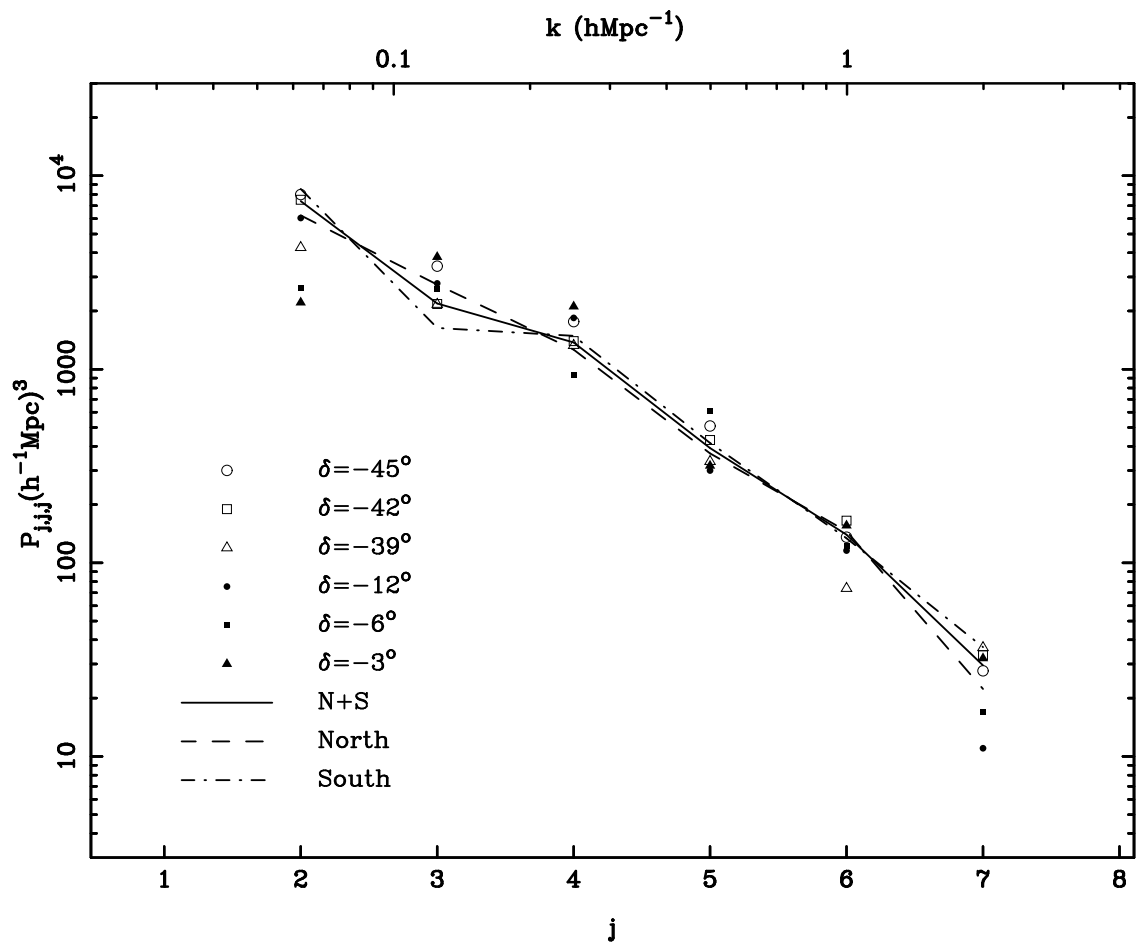}  \caption{The DWT power spectrum
measured in the flux-limited LCRS samples. The scatter symbols
represent the DWT power spectrum measured in the six slices as
indicated on the legend. The solid line show the mean power
averaged over these six slices. The dash and dot-dash line are
for the DWT power spectra calculated from the three north and
south slices enclosed in one box respectively. } \label{Fig8}
\end{center}
\end{figure}

%fig9
\begin{figure}
\begin{center}
\vspace{20.0cm} \includegraphics{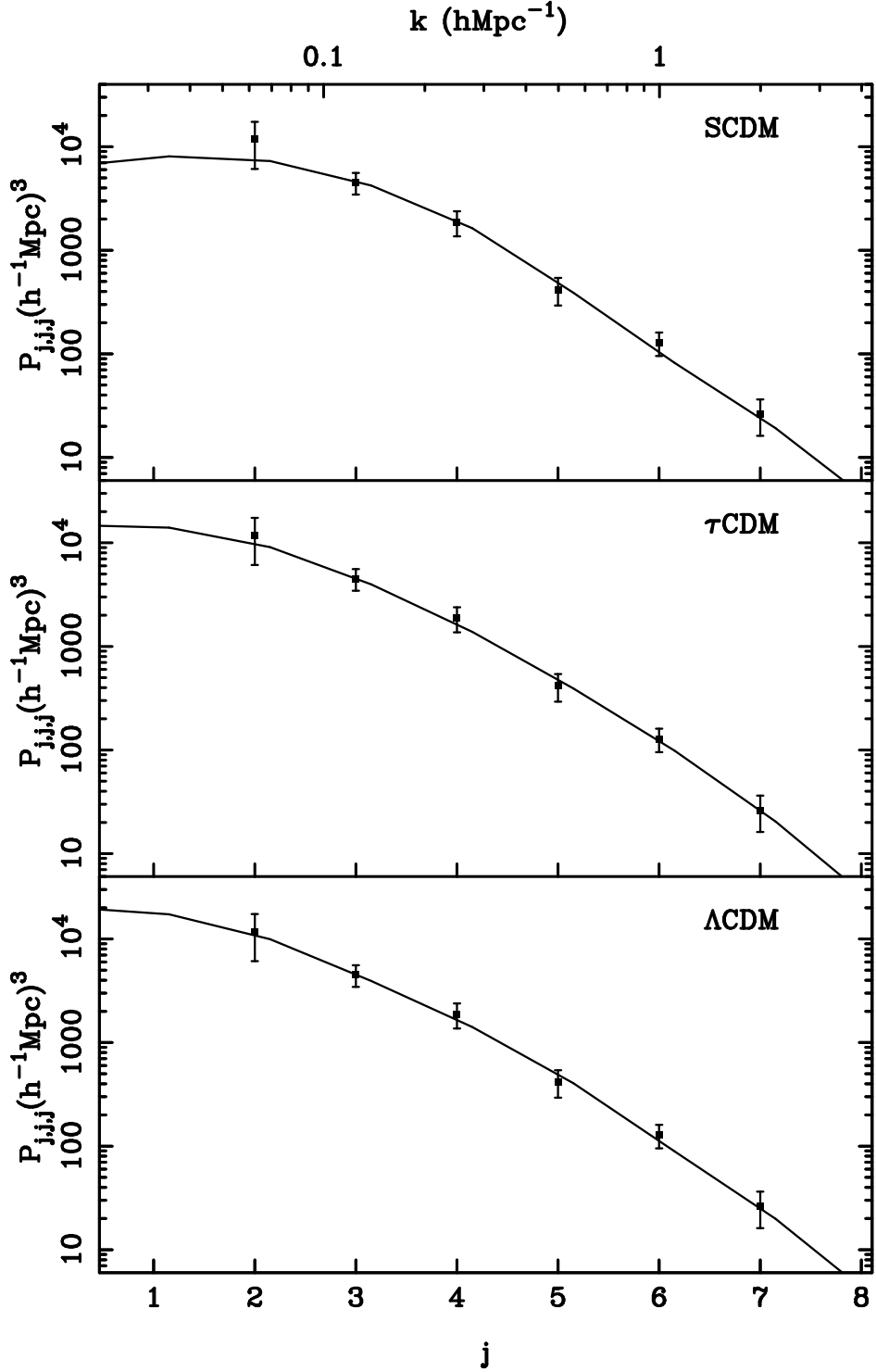} \caption{ The best-fitting (solid line) of
the LCRS diagonal DWT power spectrum for models SCDM (upper
panel), $\tau$CDM (central panel) and $\Lambda$CDM (lower panel)
with parameters listed in Table. 1. The observed values (solid
square) has been corrected for slice-like geometry effect using
the mock samples. The error bars are given by 1-$\sigma$ variance
obtained from the observed six slices.} \label{Fig9}
\end{center}
\end{figure}

%fig10
\begin{figure}
\begin{center}
\vspace{20.0cm} \includegraphics{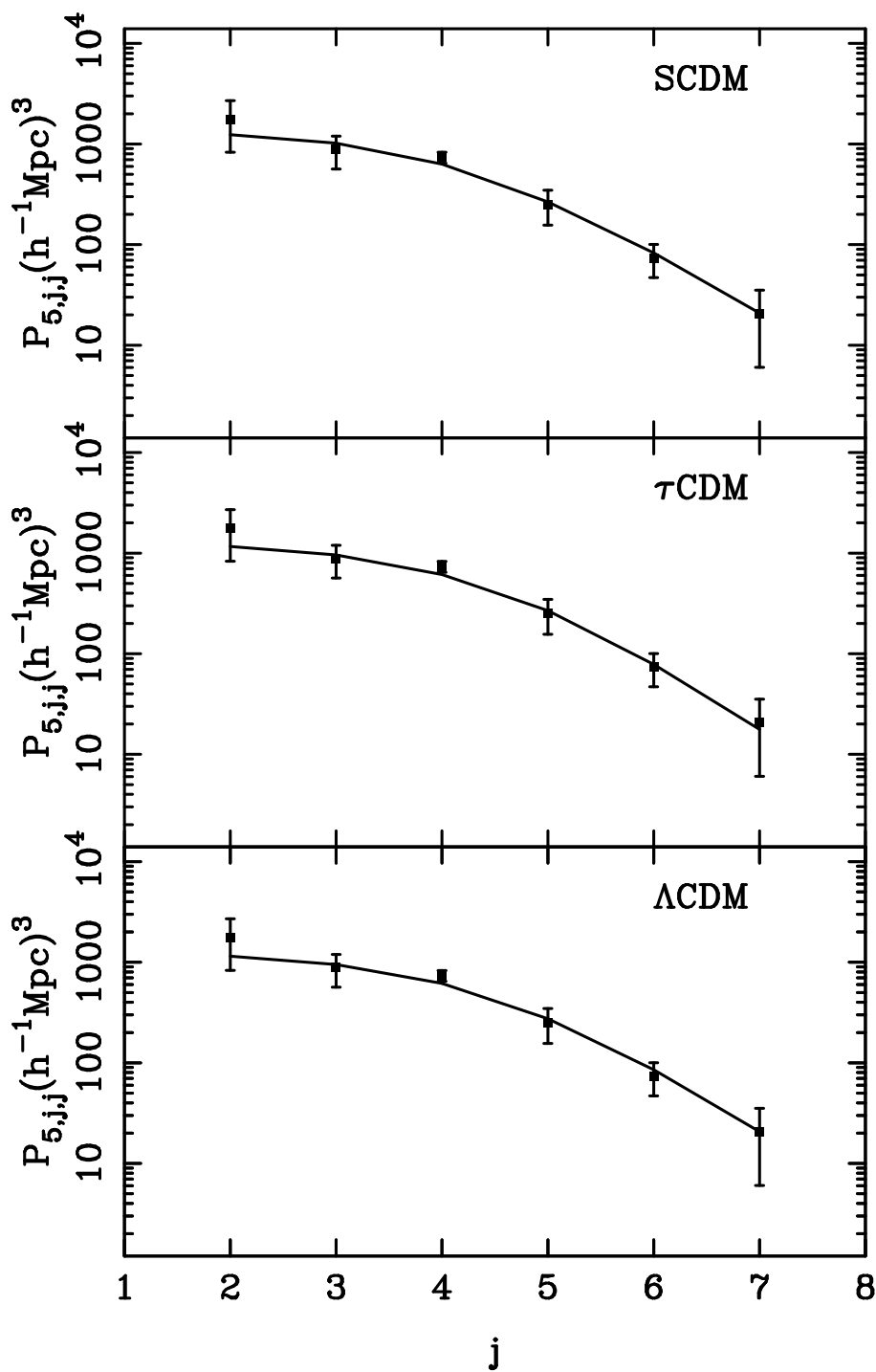} \caption{The best-fitting (solid line) of
the LCRS off-diagonal power spectrum $P_{5jj}$ for models SCDM
(upper panel), $\tau$CDM (central panel) and $\Lambda$CDM (lower
panel) with parameters listed in Table. 1. } \label{Fig10}
\end{center}
\end{figure}


\begin{thebibliography}{99}

\bibitem{bb} Bardeen, J.M., Bond, J.R., Kaiser, N. \& Szalay,
A.S., 1986, ApJ, 304, 15

\bibitem{be} Baugh, C.M., \& Efstathiou, G., 1994, MNRAS, 270, 183

\bibitem{cf} Cole, S., Fisher, K.B. \& Weinberg, D.H., 1995,
MNRAS, 275, 515

\bibitem{c} Couchman, H.M.P., 1991, ApJ, 368, 23

\bibitem{d} Daubechies I. 1992, Ten Lectures on Wavelets,
 (Philadelphia, SIAM)

\bibitem{fb} Fan, Z.H. \& Bardeen, J. M. 1995 Phys. Rev. D, 51,
6714

\bibitem{ft} Fang, L.Z. \& Thews, R. 1998, Wavelet in
Physics, (World Scientific, Singapore)

\bibitem{ff} Fang, L.Z. \& Feng, L.L. 2000, \apj, 539,5

\bibitem{f} Farge, M. Kevlahan, N., Perrier, V. \& Goirand, E. 1996,
     Proceedings of the IEEE, 84, 639

\bibitem{ff2} Feng, L.L. \& Fang, L.Z. 2000, \apj, 535,519

\bibitem{ht} Hamilton, A.J.S., Tegmark, M. \& Padmanabhan, N.,2000
astro-ph/0004334

\bibitem{j} Jing, Y.P., 1992, PhD thesis

\bibitem{jmb} Jing Y.P., Mo, H.J., \& Borner G., 1998, ApJ, 494 1

\bibitem{lsb} Landy, S.D., Szalay, A.S., \& Broadhurst, T.J.
    1998, \apj, 494, L133.

\bibitem{ls} Lin, H., Kirshner, P., Shectman, S.A., Landy, S.D.,
Oemler, A., Tucker, D.L. \& Schechter, P.L., 1996, ApJ, 464, 60L

\bibitem{pf} Pando, J. \& Fang, L.Z. 1998, Phys. Rev. E57, 3593

\bibitem{pd} Peacock, J.A. \& Dodds, S.J., 1994, MNRAS, 167, 1020

\bibitem{pd2} Peacock, J.A. \& Dodds, S.J., 1996, MNRAS, 280, L19

\bibitem{pn} Peacock, J.A., \& Nicholson, D., 1991, MNRAS, 253,
307

\bibitem{sl} Shectman, S.A., Landy, S.D., Oemler, A., Tucker,
D.L., Lin, H., Kirshner, R.P. \& Schechter, P.L., 1996, ApJ 470,
172,

\bibitem{te} Tados H. \& Efstathiou G. 1996,  MNRAS 282 1381

\bibitem{tb} Tadros H., Ballinger W. E., Taylor A. N., Heavens A.
F., Efstathiou G., Saunders W., Frenk C. S., Keeble O., McMahon
R., Maddox S. J., Oliver S., Rowan-Robinson M., Sutherland W. J.,
White S. D. M., 1999, MNRAS 305, 527

\end{thebibliography}
\end{document}